\RequirePackage{ifpdf}    
\documentclass[11pt]{article}

\pdfoutput=1
\usepackage{graphicx}
\usepackage{epsfig}
\usepackage{amsmath}
\usepackage{cite}
\usepackage{subcaption,graphicx}
\usepackage{lscape}
\usepackage{multirow}
\usepackage{slashed}
\usepackage{comment}
\usepackage{jheppub}
\usepackage{hyperref}

\usepackage{color}
\let\normalcolor\relax
\makeatletter

\newcommand{\fmslash}[2][0mu]{%
  \mathchoice
    {\fmsl@sh\displaystyle{#1}{#2}}%
    {\fmsl@sh\textstyle{#1}{#2}}%
    {\fmsl@sh\scriptstyle{#1}{#2}}%
    {\fmsl@sh\scriptscriptstyle{#1}{#2}}}
\newcommand{\fmsl@sh}[3]{%
  \m@th\ooalign{$\hfil#1\mkern#2/\hfil$\crcr$#1#3$}}
\makeatother

\newcommand{\beq}{\begin{equation}}
\newcommand{\eeq}{\end{equation}}
\newcommand{\bea}{\begin{eqnarray}}
\newcommand{\eea}{\end{eqnarray}}

\newcommand{\lsim}{{\;\raise0.3ex\hbox{$<$\kern-0.75em\raise-1.1ex\hbox{$\sim$}}\;}}
\newcommand{\gsim}{{\;\raise0.3ex\hbox{$>$\kern-0.75em\raise-1.1ex\hbox{$\sim$}}\;}}

\newcommand{\mptvec}{\not \!\! \vec{P}_T}

\title{$M_{T2}$ as a probe of CP phase in $h \rightarrow \tau \tau$ at the LHC}

\author[a,b,c]{Abhaya Kumar Swain}
\affiliation[a]{SGTB Khalsa College, University of Delhi (DU),  
	Delhi, India.}
\affiliation[b]{Department of Physics \& Astrophysics, University of Delhi,
	Delhi, India.}
\affiliation[c]{School of Physical Sciences, Indian Association for the Cultivation of Science,\\  
	2A \& B Raja S.C. Mullick Road, Jadavpur, 
	Kolkata 700 032, India.}

\emailAdd{abhayakumarswain53@gmail.com}

\abstract{We propose to utilize the transverse mass variable $M_{T2}$ and it's descendant $M_{2Cons}$ for constraining the CP admixture of the tau lepton Yukawa coupling at the LHC. We have considered the tau lepton pair produced from the Higgs boson with each tau decays to a charged pion and a neutrino, $\tau^{\pm} \rightarrow \pi^{\pm} \nu_{\tau}$. Recently, for this channel, the LHC has employed the impact parameter method to measure the CP mixing angle of tau lepton Yukawa coupling with large uncertainty. The observables we propose here can be measured in the lab frame without the impact parameter measurement and in turn, give a complementary probe of the CP admixture of tau lepton Yukawa. The CP mixing angle, with our method, can be constrained up to 17$^{\circ}$ (7$^{\circ}$) with 300 (3000) $fb^{-1}$ of integrated luminosity at the 14 LHC.}


\begin{document}
\maketitle
\section{Introduction}
\label{sec:intro}
Looking for hints of new physics beyond the Standard Model (SM) at the Large Hadron Collider (LHC) is of utmost importance after the phenomenal discovery of the Higgs boson by the CMS\cite{Chatrchyan:2012ufa} and the ATLAS\cite{:2012gk} collaborations. Measurement of the properties of the Higgs boson is an ideal place to expect new physics, if at all present, at the LHC. Many extensions of the SM modify the couplings of the SM Higgs to the gauge bosons and fermions. The CP conserving nature of the Higgs sector demands that the neutral mass eigenstates should have definite CP transformation property. Hence, measurement of the couplings of the Higgs boson to the gauge bosons and the fermions plays a key role in constraining CP violation which essentially demands new physics. With the current measurement at the LHC, the spin, parity and the couplings of the observed Higgs to gauge bosons and fermions\cite{Chatrchyan:2012jja,Aad:2013xqa,Aad:2013wqa,Chatrchyan:2013mxa,CMS:2020dkv,Sirunyan:2020sum} indicate that it is the one predicted by the SM. However, it is yet to exclude with high probability that it has a pseudoscalar admixture, in fact, the current data allows plenty of room for a pseudoscalar component\cite{CMS:2020rpr}. In this regard studying the properties of Higgs boson to third generation fermions of the SM plays a crucial role because of the larger Yukawa coupling.

Here we focus on the tau lepton Yukawa coupling measurements at the LHC. Several angular observables\cite{DellAquila:1985jin,DellAquila:1988bko,Bernreuther:1993df,Bernreuther:1993hq,Soni:1993jc,Skjold:1994qn,Grzadkowski:1995rx,Grzadkowski:1999ye,Hagiwara:2000tk,Han:2000mi,Plehn:2001nj,Choi:2002jk,Bower:2002zx,Desch:2003mw,Asakawa:2003dh,Desch:2003rw,Godbole:2004xe,Rouge:2005iy,Biswal:2005fh,Ellis:2005ika,Godbole:2007cn,BhupalDev:2007ftb,Berge:2008wi,Berge:2008dr,DeRujula:2010ys,Christensen:2010pf,Berge:2011ij,Godbole:2011hw,Harnik:2013aja,Berge:2013jra,Chen:2014ona,Dolan:2014upa,Hayreter:2015cia,Han:2016bvf} have been proposed already in the literature to measure the CP-even and CP-odd mixing angle of the tau lepton Yukawa coupling. However, most of them are defined in the CM frame of the Higgs boson or tau lepton rest frame which is extremely challenging to reconstruct. This is mainly because there are two (at least) neutrinos in the final state which escape detection and, in addition, the center of mass energy of collision between the partons at the LHC is unknown. To circumvent these difficulties additional independent measurements are required. Fortunately, the tau lepton has finite decay length which renders measurable impact parameter at the LHC which can then be used for constructing CP observables\cite{Berge:2008dr,Berge:2014sra} in the charged particles CM frame (same as tau lepton pair CM frame). The measured impact parameter can also be utilized to reconstruct tau lepton momenta\cite{Hagiwara:2016zqz,Bhardwaj:2016lcu} which can be used to define the CP sensitive observables in the CM frame of tau lepton pair.

Recently, the CMS collaboration\cite{CMS:2020rpr} has measured the coupling of the Higgs to tau lepton pair with the center of mass energy of 13 TeV for 137 fb$^{-1}$ of integrated luminosity. They have considered one of the two taus decays to muon with the other decays hadronically and also considered both the tau decay hadronically which accounts for 50$\%$ branching fraction of total Higgs to tau lepton pair. The decay planes of tau leptons produced from the Higgs exhibit angular correlations which is utilized to put a limit on the CP-odd component of the tau Yukawa coupling. The impact parameter vector is utilized as an additional measurement when the tau decay product involves a single charged particle in the final state to define the decay plane while the accompanying neutral pion is used otherwise. With this, the mixing angle between CP-even and CP-odd component is constrained to be $4^{\circ} \pm 17^{\circ}$ with large uncertainty of $\pm 36^{\circ}$ at 95$\%$ confidence level. Among all the above tau decay channels they considered, the most sensitive channel is the one where, at least, one of the two taus decays hadronically via $\rho$ meson. The process with each tau decays to a single charged hadron (muon) and a neutrino(s) is comparatively less sensitive because the impact parameter vector has relatively large uncertainties in CMS.

In this article we focus on the tau lepton pair produced from the Higgs boson with each tau decays to a single charged pion and a neutrino, $\tau^{\pm} \rightarrow \pi^{\pm} \nu_{\tau}$, and propose a lab frame observable to constrain the CP admixture of the tau Yukawa coupling. We advocate to utilize the transverse mass variable, $M_{T2}$\cite{Lester:1999tx, Barr:2003rg, Meade:2006dw, Lester:2007fq, Cho:2007qv, Cho:2007dh, Barr:2007hy, Gripaios:2007is, Nojiri:2008hy,Burns:2008va,Konar:2009wn,Cho:2009ve, Cho:2010vz,Barr:2009jv,Tovey:2008ui, Polesello:2009rn, Serna:2008zk,Matchev:2009ad}, which efficiently minimizes the neutrino momenta satisfying the missing transverse energy constraints and predicts the tau lepton mass. The kinematic variable $M_{T2}$ shows singular behavior\cite{Kim:2009si,Rujula:2011qn,Matchev:2019bon,Park:2020rol} which is endpoint in its distribution and it is at the tau lepton mass. This variable not only predicts the mass but also measures the CP mixing angle of tau lepton Yukawa coupling. We discuss this variable briefly in the following sections. We then analyze the effect of parton showering on $M_{T2}$ to realize that the CP sensitivity of it reduces significantly. The transverse boost resulting from the parton showering is held responsible for this because $M_{T2}$ is not, in general, a boost invariant quantity though it remains invariant under longitudinal boost. Fortunately, the (1+3)$-$dimensional generalization of $M_{T2}$, the constrained mass variable $M_{2Cons}$\cite{Konar:2015hea,Konar:2016wbh}, is turned out to be extremely useful here because it remains invariant under both the transverse and longitudinal boost owing to its construction in (1+3)$-$dimension. We discuss this variable briefly in the following sections and construct asymmetry for the constrained mass variable $M_{2Cons}$ to measure the CP-even and CP-odd mixing angle. 

As mentioned above we have considered the process $p p \rightarrow H \rightarrow \tau^{+}\tau^{-} \rightarrow \pi^{+}\pi^{-}\bar{\nu}_{\tau}\nu_{\tau}$ at the LHC. The lagrangion in terms of CP-even and CP-odd mixing angle of tau lepton Yukawa coupling which serves our purpose is,
\begin{equation}\label{Lagrangian}
\mathcal L \supset - \frac{m_\tau}{v}H\bar \tau (\cos\theta_{\tau} + i\gamma_5\sin\theta_{\tau})\tau,
\end{equation}
where $m_{\tau}$ is the tau lepton mass and v is the vacuum expectation value which has a value of 246 GeV. The CP mixing angle of tau Yukawa coupling is denoted by $\theta_{\tau}$.

The rest of the paper is organized as follows, in section~\ref{sec:MT2} we briefly discuss the transverse mass $M_{T2}$ and its (1+3)$-$dimensional generalization, the constrained mass variable $M_{2Cons}$. In section~\ref{sec:shape} the shape of both the variables $M_{T2}$ and $M_{2Cons}$ are discussed including the impact of transverse recoil from parton showering on these variables. Section~\ref{sec:results} discuss the sensitivity of the variable to constrain the CP mixing angle of the tau lepton Yukawa coupling and followed by summary and conclusion in section~\ref{sec:conclusion}.


\section{Stransverse mass variable and its 3D generalization}
\label{sec:MT2}
The transverse mass variable, $M_{T2}$ is a widely used kinematic observable both in phenomenological and experimental studies in events associated with missing transverse energy. Although it was initially proposed for the determination of the mass of parent and/or daughter particles, later on, it was realized that $M_{T2}$ can play an extremely important role in excluding/discovering new physics at the collider experiments. Moreover, $M_{T2}$ can also be used for measuring the properties like spin and reconstructing invisible particle momentum, etc. of the new physics after its discovery. In this section, we briefly discuss $M_{T2}$ and the variable $M_{2Cons}$ which is a 3-dimensional generalization of $M_{T2}$ with available kinematic constraints in each event. The transverse mass variable $M_{T2}$ is defined as the maximum transverse mass between pair of the parents following a minimization with respect to invisible momenta satisfying the missing transverse momentum constraints,
\begin{equation}\label{MT2}
M_{T2}     \equiv     \min_{\substack{\vec{q}_{iT} \\ \{\sum \vec{q}_{iT} = \mptvec \}}} 
\left[  \max_{i =1, 2}      \{  M_T^{(i)}  ({p}_{iT}, {q}_{iT}, m_{vis(i)}; m_{\nu_{\tau}})  \}  \right].
\end{equation}
Where the definition of the transverse mass for each parent is,
\begin{eqnarray}
&&(M_T^{(i)})^2 = m_{vis(i)}^2 + m_{\nu_{\tau}}^2 + 2(E_T^{vis(i)}E_T^{inv(i)} - \vec{p}_{iT}.\vec{q}_{iT})\\  \label{mt}
&&E_T^{vis(i)} = \sqrt{m_{vis(i)}^2 + p_{iT}^2}, \, \,\,\,E_T^{inv(i)} = \sqrt{m_{\nu_{\tau}}^2 + q_{iT}^2}  \, \, .
\end{eqnarray}

The $E_T^{vis(i)}$ and $E_T^{inv(i)}$ are the transverse energy of the visible and invisible particles respectively. And the quantities $\vec{p}_{iT}$ and $\vec{q}_{iT}$ are the transverse momenta of the visible and invisible particles respectively. We note here that in this article the visible and invisible particle corresponds to the charged pion and the tau neutrino produced from tau lepton decay. The masses of pions and neutrinos are denoted as $m_{vis(i)}$ and $m_{\nu_{\tau}}$. By construction, the variable $M_{T2}$ is bounded by the parent mass for the correct invisible particle mass. In this analysis the invisible particle is the neutrino, so we can neglect its mass, so $M_{T2}$ is bounded from above by the mass of tau-lepton, $m_{\tau}$.

Clearly, the transverse mass variable $M_{T2}$, as shown in eqn.~\ref{MT2}, is not utilizing the available longitudinal momentum of the visible particles. The longitudinal component also carries important information and any variable which involves these including the transverse component might gain some advantage compared to the $M_{T2}$. For example, the inclusion of the longitudinal component information enables one to make use of the mass-shell constraints, if available, which are $(1+3)$-dimensional. Hence, a $(1+3)$-dimensional generalization of $M_{T2}$ is defined and dubbed as $M_2$ with all the above mentioned capabilities and the details can be found in \cite{Mahbubani:2012kx, Cho:2014naa,Kim:2017awi,Debnath:2017ktz}. In this paper, we discuss a  $(1+3)$-dimensional mass variables with the Higgs mass-shell constraint including the missing transverse momentum constraints which was originally proposed in~\cite{Konar:2015hea} and defined as,
\begin{equation}\label{M2Cons}
M_{2Cons}
\equiv  \min_{\substack{\vec{q}_{1}, \vec{q}_{2} \\   
		\left\{   \substack{  \vec{q}_{1T} + \vec{q}_{2T} = {\mptvec} \\  (p_1 + p_2 + q_1+ q_2)^2 = m_H^2 }  \right\}   }} 
\left[ \max_{i =1, 2}  \{ M^{(i)}(p_{i}, q_{i}, m_{vis(i)}; m_{\nu_{\tau}}) \} \right],
\end{equation}
where $(M^{(i)})^2 = m_{vis(i)}^2 + m_{\nu_{\tau}}^2 + 2(E^{vis(i)}E^{inv(i)} - \vec{p}_{i}.\vec{q}_{i})$. The quantities $\{E^{vis(i)}, \vec{p}_{i}\}$ and  $\{E^{inv(i)}, \vec{q}_{i}\}$ are the visible and invisible particles four momenta respectively. $m_H$ is the mass of the Higgs boson and is used as a constraint. The variable $M_{2Cons}$ inherits all the properties of the $M_{T2}$ like bounded from above by the parent mass, $m_{\tau}$. In addition, it predicts a higher value compared to $M_{T2}$ event by event and there by increasing the number of events at the end-point. More intrestingly, it can measure both the parent and daughter mass, for a process with massive daughter, simultaneously unlike $M_{T2}$ which gives a relation between them in the absence of extra transverse momentum. Moreover,  $M_{2Cons}$ also predicts a better momentum approximation of the invisible particles.
\section{Analysing shape of the $M_{T2}$ and $M_{2Cons}$}
\label{sec:shape}
\begin{figure}[t]
	\centering
	\includegraphics[scale=0.6,keepaspectratio=true,  angle=0]{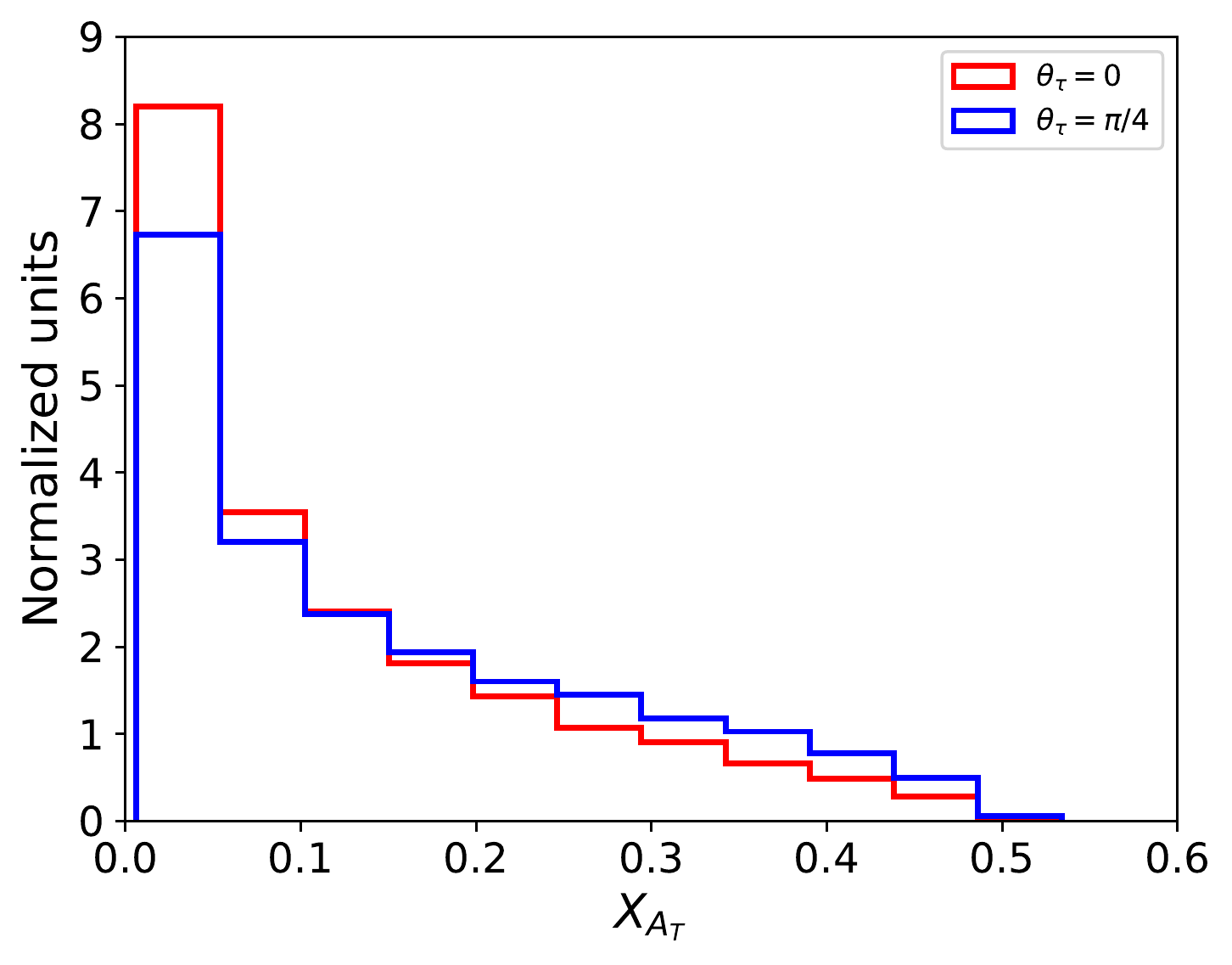}
	\caption{The distribution of the dimensionless quantity $X_{A_T}$ is displayed. The red colored histogram is for the CP phase $\theta_{\tau} = 0$ while the histogram in blue color is for $\theta_{\tau} = \pi/4$.}
	\label{fig:ATDistribution}
\end{figure}
Before we discuss the shape of the $M_{T2}$, we briefly mention the analytical formula of the transverse mass variable. As discussed in the introduction we consider the one prong decay of tau-lepton with only a single visible particle, charged pion, and a neutrino. So the analytical formula is driven by a balanced momentum configuration known as balanced solution~\cite{Cho:2007dh} as,
\begin{equation}\label{MT2Analytical}
M_{T2} = \frac{1}{\sqrt{2}}(\sqrt{A_T + m_{\pi}^2} + \sqrt{A_T - m_{\pi}^2}).
\end{equation}
With
\begin{equation}\label{AT}
A_T = E_T^{v1} E_T^{v2} + |\vec{p}_{1T}||\vec{p}_{2T}|\cos\alpha,
\end{equation}
where $\alpha$ is the angle between the charged pions in the transverse direction and $m_{\pi}$ is pion mass. In eqn.~\ref{MT2Analytical} and in the rest of the paper, we have assumed the mass of tau neutrino is zero. It is clear from eqn.~\ref{AT} that the shape of the transverse mass variable, $M_{T2}$, is governed by the orientation of the charged pions in the transverse direction. For example, the minimum of the $M_{T2}$ is the pion mass and it occurs when the two pions are back to back in the transverse direction. Since both the tau-leptons from Higgs are highly boosted, the charged pions travel mostly along the tau direction as a result the pions are widely separated. Hence, the maximum of $M_{T2}$ occurs when two pions are slightly away from the back to back momentum configuration. Interestingly, the orientation of the charged pions in the transverse direction is sensitive to the CP phase of the tau lepton Yukawa coupling as evident in fig.~\ref{fig:ATDistribution}. Where we have displayed the distribution of a dimensionless quantity $X_{A_T} = \frac{A_T}{m_{\tau}^2}$. The red colored histogram is for the CP phase $\theta_{\tau} = 0$ and the blue colored one is for the CP phase $\theta_{\tau} = \pi/4$ respectively. Moreover, there are relatively more events in back to back charged pions for $\theta_{\tau} = 0$ compared to $\theta_{\tau} = \pi/4$ which is undoubtedly visible from the distribution. Hence, the major source of the CP phase in the transverse mass variable $M_{T2}$ is the orientation of the charged pions in the transverse direction.

Similarly, we define another two dimensionless ratio for the transverse mass variable $M_{T2}$ and for the variable $M_{2Cons}$ as,
\begin{equation}\label{XMT2M2Cons}
X_{\beta} = \frac{\beta}{m_{\tau}},
\end{equation}
with $\beta = \{M_{T2}, M_{2Cons}\}$. Since both the variables $M_{T2}$ and $M_{2Cons}$ are bounded by the tau-lepton mass, the dimension less quantity $X_{\beta}$ varies between 0 and 1.
\begin{figure}[t]
	\centering
	\includegraphics[scale=0.51,keepaspectratio=true,  angle=0]{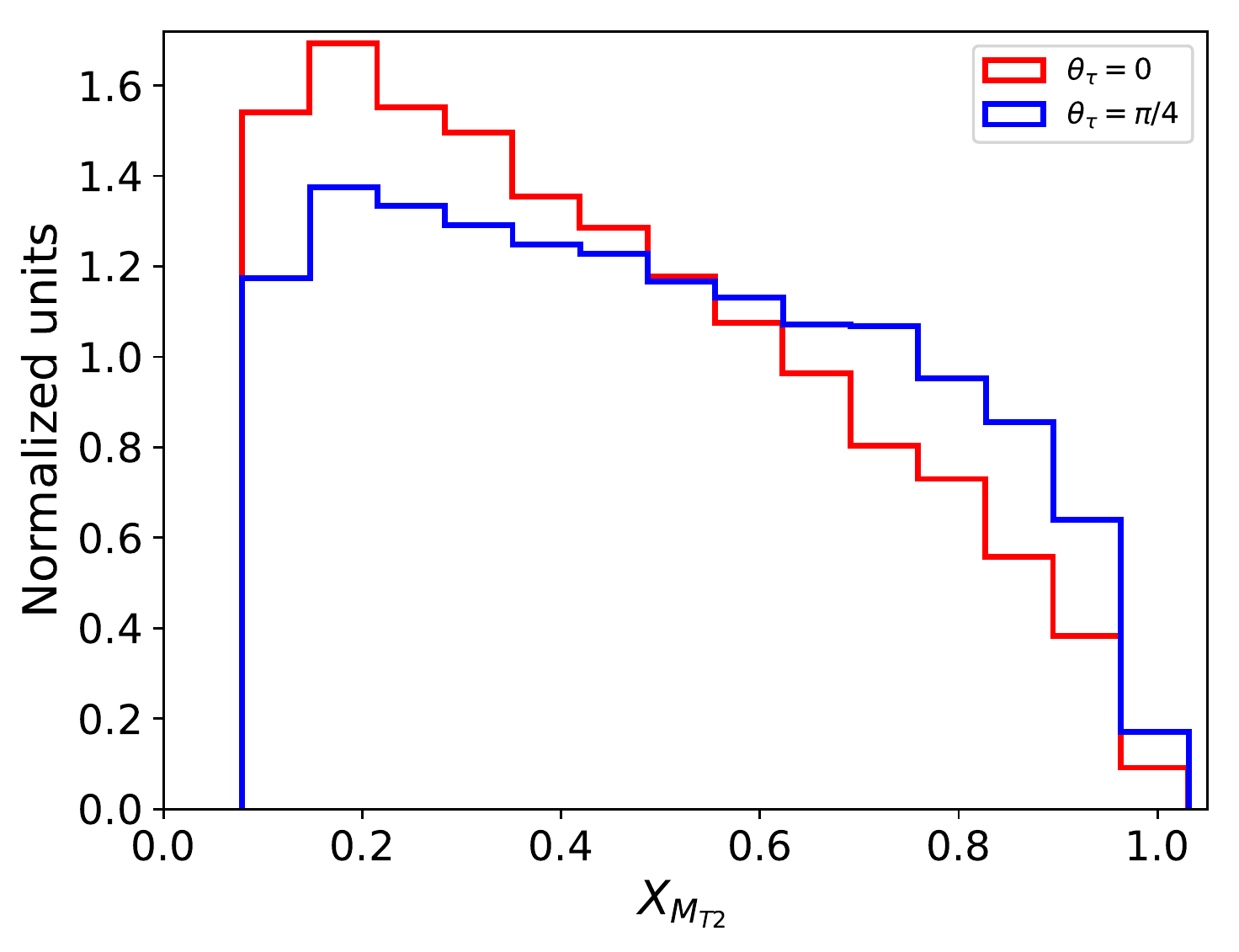}
	\includegraphics[scale=0.51,keepaspectratio=true,  angle=0]{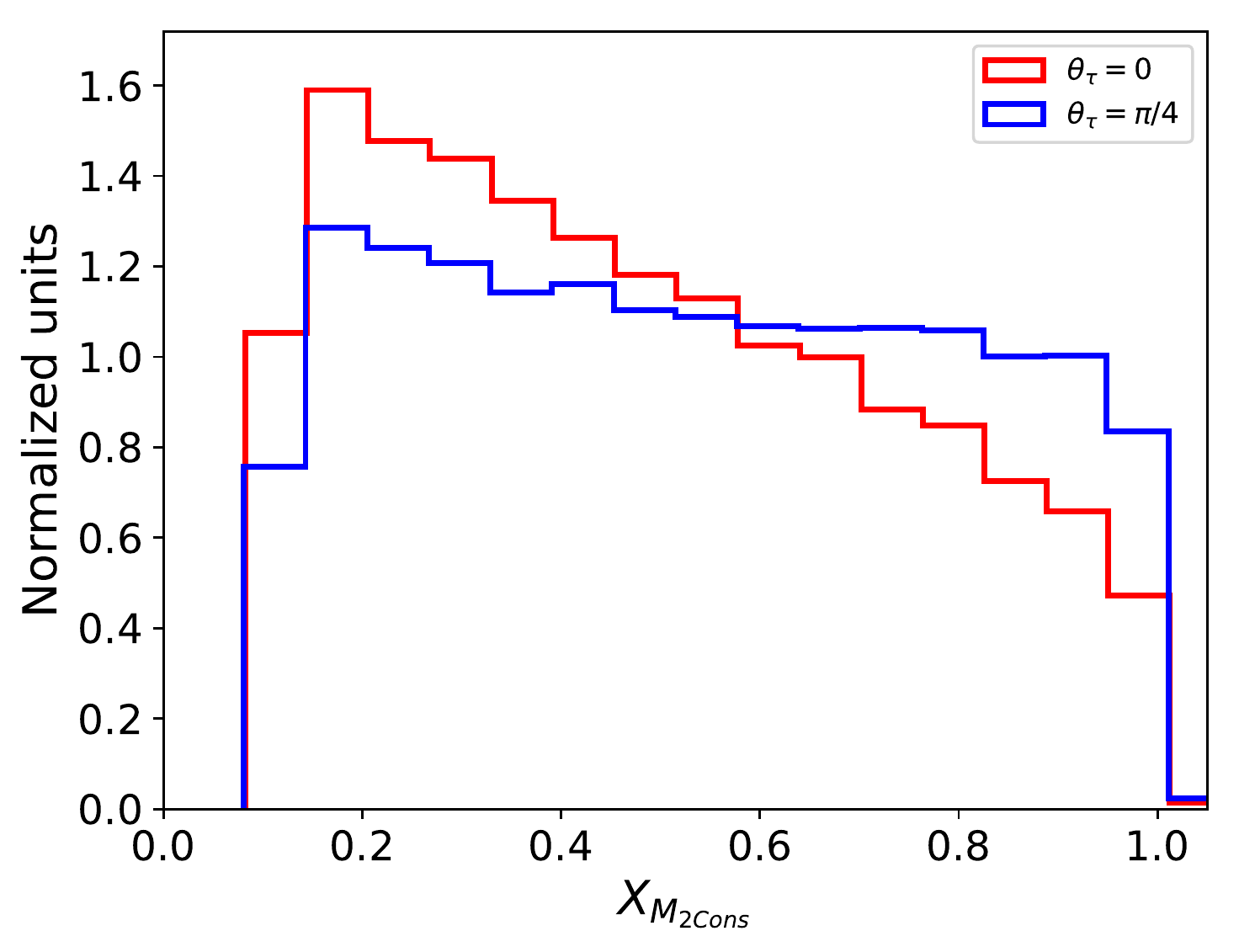}
	\caption{Normalized distribution for the quantity $X_{M_{T2}}$ is delineated in the left panel while the right panel portrays  $X_{M_{2Cons}}$. For both the panel the blue colored histograms represent the CP phase $\theta_{\tau} = 0$ case and the red distributions correspond to $\theta_{\tau} = \pi/4$. Undoubtedly both the observables distinguish the  $\theta_{\tau} = 0$ case from  $\theta_{\tau}= \pi/4$ quite well.}
	\label{fig:MT2M2ConsDistribution}
\end{figure}
The distribution for the variables  $M_{T2}$ (left panel) and $M_{2Cons}$ (right panel) are displayed in fig.~\ref{fig:MT2M2ConsDistribution} with blue colored histograms correspond to  $\theta_{\tau} = 0$ while the red colored histogram is for the CP phase $\theta_{\tau} = \pi/4$ case. As discussed earlier, the variable $M_{T2}$ incorporates the CP phase information in its shape which clearly visible from the figure. Hence, the shape of $M_{T2}$ can be used to constrain the CP phase which will be discussed in section~\ref{sec:results}. Since $M_{2Cons}$ is the 3-dimensional generalization of $M_{T2}$ with the advantages of utilizing the Higgs mass shell constraint, the number of events towards the endpoint increases compared to earlier observable which in turn result in improved mass measurement. Moreover, $M_{2Cons}$ provides a better approximation for the invisible particles, neutrinos, momentum compared to $M_{T2}$, details about this variable can be found in \cite{Konar:2015hea}. The CP sensitivity of $M_{2Cons}$ is similar to $M_{T2}$ which is evident from the right panel fig.~\ref{fig:MT2M2ConsDistribution}.

Since the transverse mass variable $M_{T2}$ as shown in eqn.~\ref{MT2} involves only the transverse momenta, it is not in general transverse boost invariant~\cite{Tovey:2019znm} though it is longitudinal boost invariant. In addition, these observables are expected to be measured at the LHC where additional jets maybe there which might come from parton showering and/or additional jets produced at the matrix element. Although these extra jets resulting from parton showering are very helpful in mass measurement by displaying a kink~\cite{Cho:2007qv,Gripaios:2007is,Barr:2007hy,Cho:2007dh} in $M_{T2}$ endpoint, but will change the orientation of the charged pions and hence the shape of the $M_{T2}$ is affected significantly. Here the mass bound variable $M_{2Cons}$ comes as a cure owing to its dimensionality, it is not only longitudinal boost invariant but also transverse boost invariant which makes it immune to the effects of parton showering and additional jet effects. 
\begin{figure}[t]
	\centering
	\includegraphics[scale=0.52,keepaspectratio=true,  angle=0]{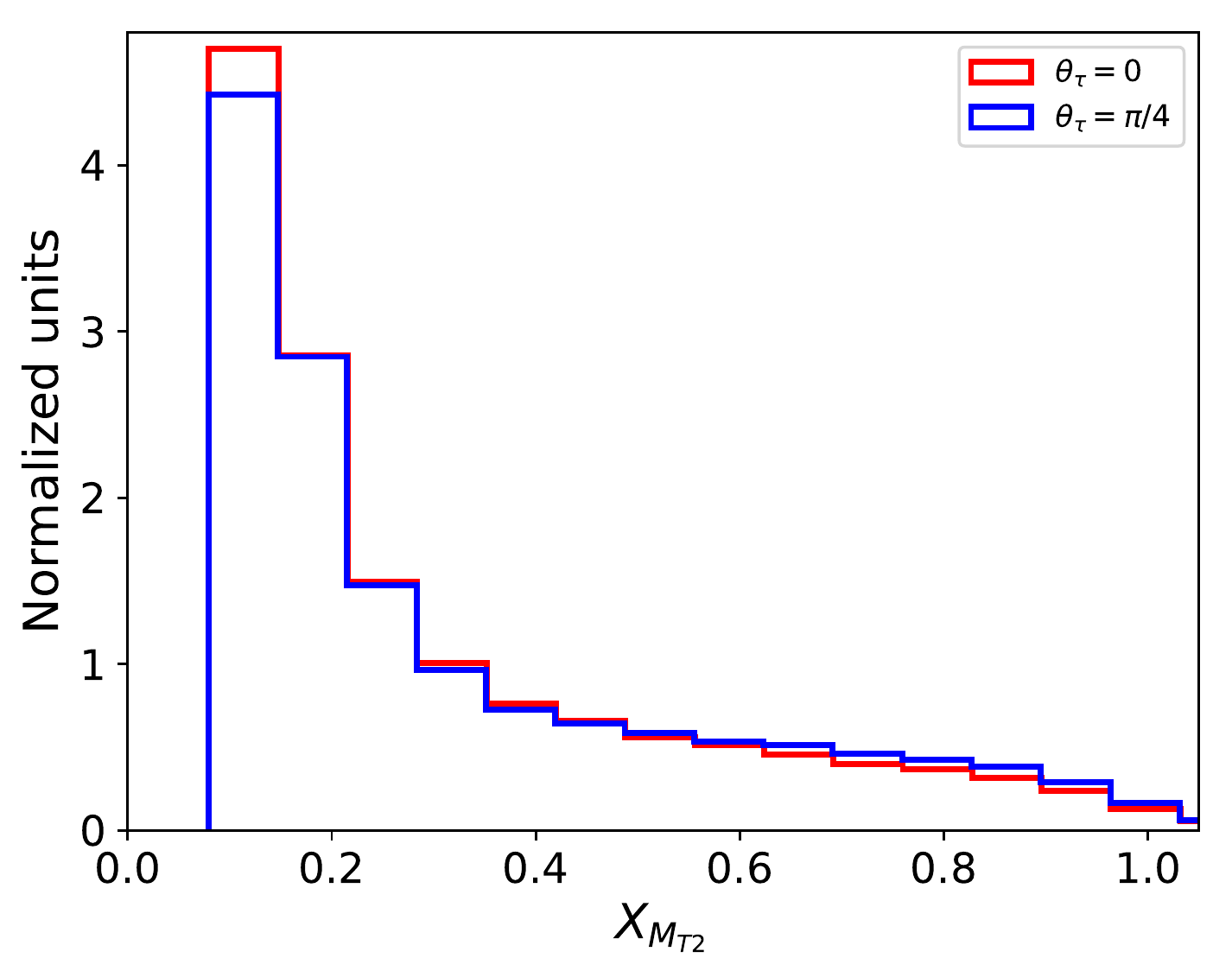}
	\includegraphics[scale=0.52,keepaspectratio=true,  angle=0]{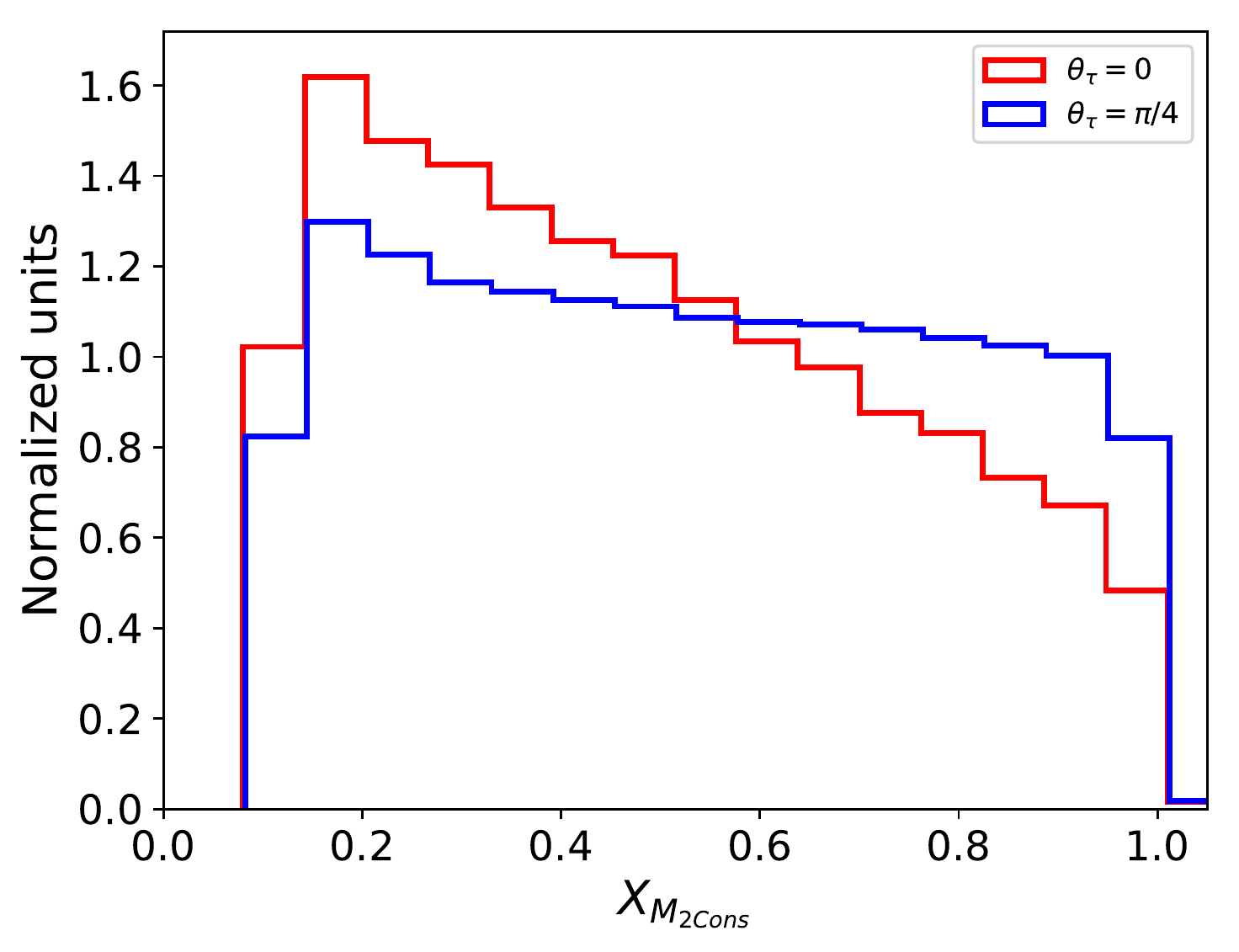}
	\caption{Effect of additional boost resulting from the parton showering on the variable $M_{T2}$ and $M_{2Cons}$ is displayed. The normlized distribution for $X_{M_{T2}}$ (left panel) and for $X_{M_{2Cons}}$ (right panel) is portrayed with blue colored histogram corresponds to the SM case, $\theta_{\tau} = 0$, and the red colored distribution is for $\theta_{\tau} = \pi/4$. Evidently the shape of the $M_{T2}$ is modified significantly because of the transverse boost resulting from the parton showering. However, the variable $M_{2Cons}$ is not much affected.}
	\label{fig:MT2M2ConsWithPythiaON}
\end{figure}
These effects are displayed in fig.~\ref{fig:MT2M2ConsWithPythiaON} with the variable $M_{T2}$ distribution in the left panel while the distribution of $M_{2Cons}$ on the right panel. Undoubtedly the shape the variable $M_{T2}$ is changed significantly because of the transverse boost coming from the parton showering. But the variable $M_{2Cons}$, as expected, is transverse boost invariant as a result its distribution remains unaffected after parton showering. 

These kinematic variables are calculated using the package\footnote{Recently, another interesting library for calculating constrained mass variables known as {\it YAM2}~\cite{Park:2020bsu} is developed which can also be used for obtaining kinematic variables utilized here.} {\it Optimass}\cite{Cho:2015laa} and also verified by independent codes written in Mathematica. We have simulated parton level events using {\it Madgraph5}\cite{Alwall:2014hca} with model file from Higgs Characterisation model\cite{Artoisenet:2013puc} written in {\it Feynrules}\cite{Alloul:2013bka}. After the parton level events, the showering is performed using {\it Pythia8}\cite{Sjostrand:2007gs}. We have performed analysis that accounts for a detailed detector level analysis for observables discussed here in a companion paper\cite{Abhayaprep:2020abc}. 

\section{Results and Discussion}
\label{sec:results}
Here we discuss the sensitivity of the variable $M_{2Cons}$ to constrain the CP phase. Since the variable $M_{T2}$ is not remaining much sensitive to the CP phase after parton showering, we refrain ourselves to give its sensitivity plot. We have constructed asymmetry corresponding to the quantity $X_{M_{2Cons}}$ which is defined as,
\begin{equation}\label{XM2Cons_sensitivity}
A_{M_{2Cons}} = \frac{\mathcal{N}(X_{M_{2Cons}} > 0.5) - \mathcal{N}(X_{M_{2Cons}} < 0.5)}{\mathcal{N}(X_{M_{2Cons}} > 0.5) + \mathcal{N}(X_{M_{2Cons}} < 0.5)},
\end{equation}
where $\mathcal{N}$ is the number of events. The sensitivity is judged based on the variation of asymmetry with respect to the CP phase $\theta_{\tau}$. Here we have only included the parton shower effects and have not considered the hadronization, detector effects, and also the backgrounds. Those effects on $M_{2Cons}$ are analyzed in a companion paper\cite{Abhayaprep:2020abc}.
\begin{figure}[t]
	\centering
	\includegraphics[scale=0.47,keepaspectratio=true,  angle=0]{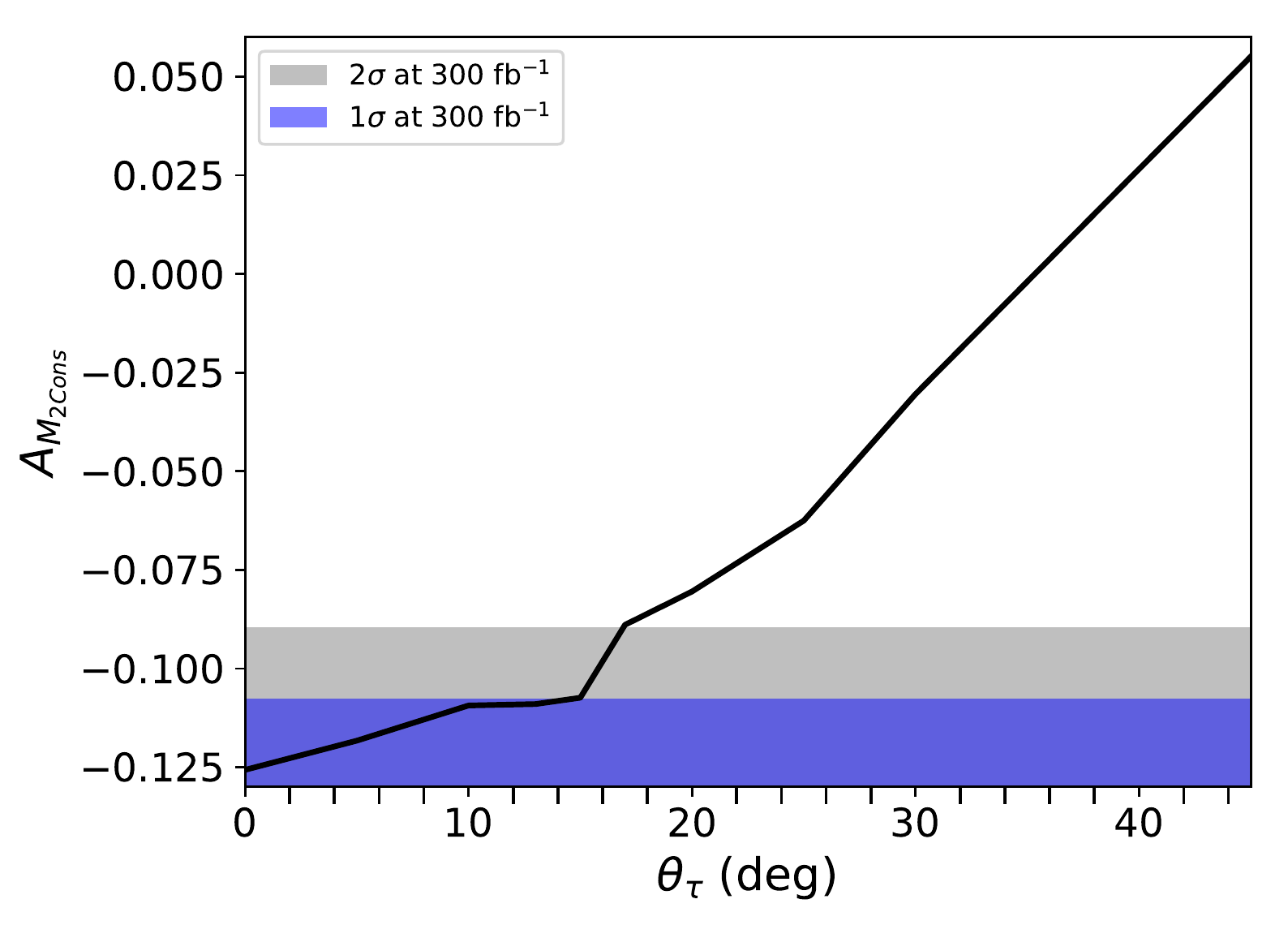}
	\includegraphics[scale=0.47,keepaspectratio=true,  angle=0]{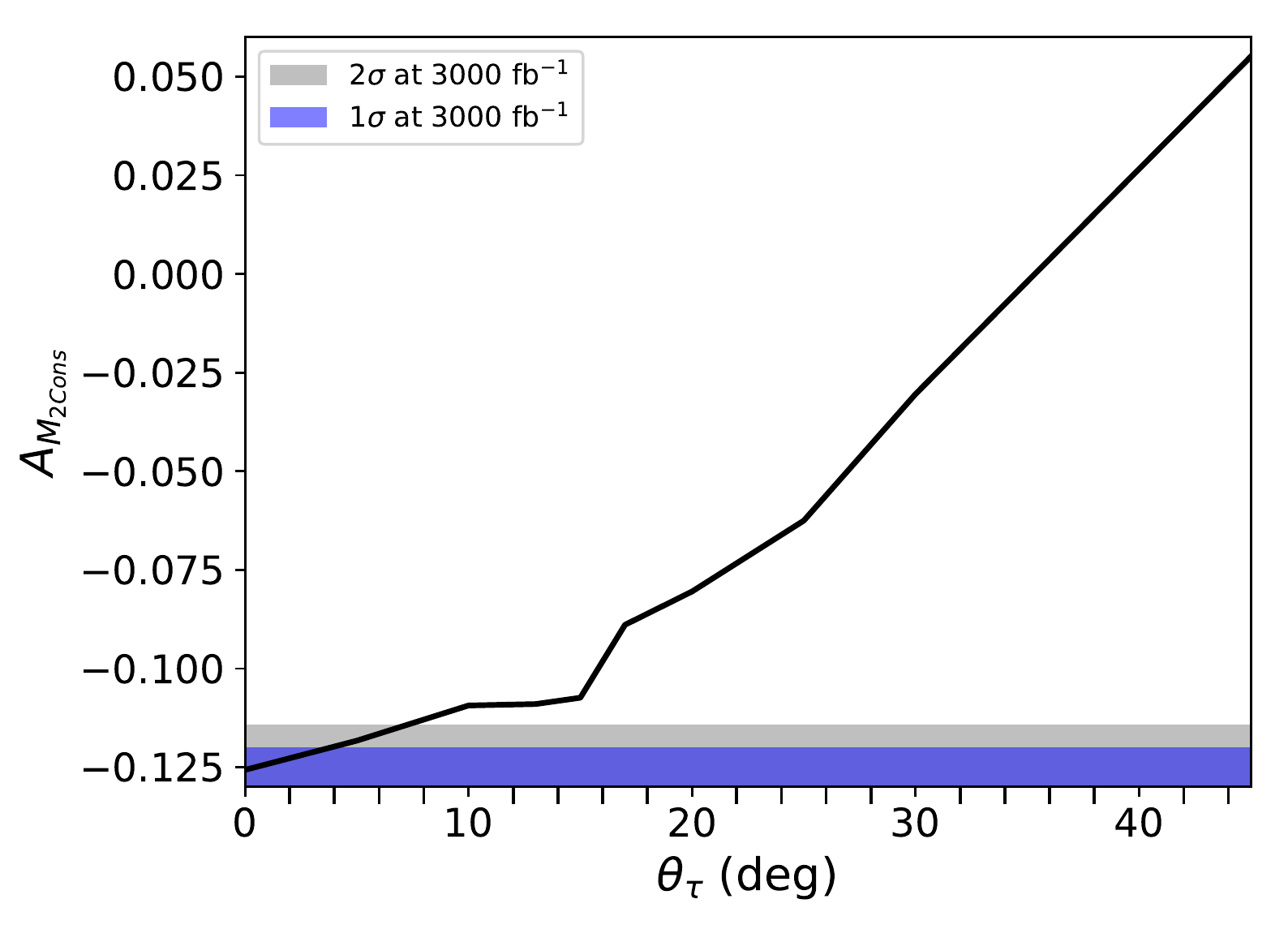}
	\caption{Asymmetry for the quantity $X_{M_{2Cons}}$ is represented by the black solid line. The blue and the gray band correspond to the 1$\sigma$ and the 2$\sigma$ uncertainty at the 300 (left panel) and 3000 $fb^{-1}$ (right panel) integrated luminosity at 14 TeV LHC. In order to achieve a reliable estimate of the sensitivity for the CP phase, we have included the realistic tau reconstruction and identification efficiency in the uncertainty calculation.}
	\label{fig:XM2ConsSensitivity}
\end{figure}
The statistical uncertainty corresponding to $A_{M_{2Cons}}$ is given by,
\begin{equation}\label{uncertainty}
\Delta \mathcal A=\frac{\sqrt{1-\mathcal (A_{M_{2Cons}}^{\rm SM})^2}}{\sqrt{\sigma_{\rm SM} ~\epsilon~ {\mathcal L} }},
\end{equation}
where the ${\mathcal L}$ is the integrated luminosity. To get a reliable estimate of the sensitivity we have included an efficiency factor for detecting di$-$tau events in the realistic LHC environment and also after the elimination of backgrounds. In eqn.~\ref{uncertainty} $\epsilon$ corresponds to this efficiency for both the tau-lepton which is 0.302. Here we have taken combined tau reconstruction and identification efficiency as 0.55~\cite{TheATLAScollaboration:2015lks} with a Medium working point. The asymmetry $A_{M_{2Cons}}^{\rm SM}$ corresponds to the SM tau pair events produced from the Higgs boson with $\sigma_{\rm SM}$ is its production cross section which is 55.22 pb~\cite{Cepeda:2019klc} for gluon fusion at the 14 TeV LHC. In fig.~\ref{fig:XM2ConsSensitivity} the asymmetry for $X_{M_{2Cons}}$ is displayed in solid black line. The blue and the gray band correspond to the 1$\sigma$ and the 2$\sigma$ statistical uncertainty for the 14 TeV LHC at 300 $fb^{-1}$ of integrated luminosity for the left panel while for the right panel the uncertainties are displayed for 3000 $fb^{-1}$ of integrated luminosity. Using the lab frame kinematic variable $X_{M_{2Cons}}$ we can constrain the CP phase $\theta_{\tau}$ upto 17$^{\circ}$ (7$^{\circ}$) with 300 (3000) $fb^{-1}$ of integrated luminosity at 14 LHC.
\section{Summary and Conclusion}
\label{sec:conclusion}
After the discovery of the Higgs boson at the LHC measurement of its properties is of supreme importance not only to confirm that it is indeed the SM Higgs boson but also it is an ideal window to search for new physics beyond the SM. Although the current measurement of the properties like the spin, parity, and couplings of the Higgs boson at the LHC indicate that it is the one predicted by the SM but there is enough room to allow for a CP-odd component in the tau Yukawa coupling. Hence, measuring the CP mixing angle of the tau lepton Yukawa coupling is extremely important. In this article, we propose to utilize the transverse mass variable $M_{T2}$ and its successor $M_{2Cons}$ for constraining the CP admixture of tau Yukawa coupling. Specifically, we advocate the use of $M_{2Cons}$ because of its advantages that it is both longitudinal and transverse boost invariant which makes it sensitive to the CP phase in the lab frame at the LHC. The CP mixing angle of the tau lepton Yukawa coupling using $M_{2Cons}$ can be constrained up to 17$^{\circ}$ (7$^{\circ}$) with 300 (3000) $fb^{-1}$ of integrated luminosity at the 14 LHC.

\bigskip
\acknowledgments
This work was supported by Department of Physics, SGTB Khalsa College, University of Delhi in the SERB sponsored Multi Institutional project titled "Probing New Physics Interactions" (CRG/2018/004889). We thank Partha Konar, Pankaj Sharma, Sukanta Dutta, Ashok Goyal, Mukesh Kumar and Kai Ma for useful discussion. We thank Tanmoy Mondal, Urmila Das, Akanksha Bhardwaj and Disha Bhatia for going through the manuscript and pointing out important suggestions to improve it. We also thank the Department of Atomic Energy, Government of India, for the Regional Centre for Accelerator-based Particle Physics (RECAPP), Harish-Chandra Research Institute (HRI) for hospitality where initial idea was discussed.


\bibliographystyle{unsrt}
\bibliography{bibliography}

\end{document}